\documentclass[
pra, reprint, twocolumn, nofootinbib]{revtex4-2}

\usepackage{times}

\usepackage{fancyhdr}
\usepackage{graphicx}
\usepackage{float}
\usepackage{tabularx,booktabs} 
\usepackage{enumitem}

\usepackage{physics}
\usepackage{amsmath}
\usepackage{amssymb}
\usepackage{dsfont}
\usepackage{mathrsfs} 
\usepackage{euscript}
\usepackage{upgreek}
\usepackage{mathtools}
\usepackage{mathdots}
\usepackage[bbgreekl]{mathbbol}
\DeclareSymbolFontAlphabet{\mathbbol}{bbold}
\DeclareSymbolFontAlphabet{\mathbb}{AMSb}

\usepackage{amsmath}
\usepackage{amsthm}
\theoremstyle{remark}

\makeatletter
\def\maketag@@@#1{\hbox{\m@th\normalfont\normalsize#1}}  
\makeatother

\usepackage[usenames,dvipsnames,table]{xcolor}
\definecolor{blue_ref}{RGB}{46,48,146}

\usepackage[colorlinks=true,urlcolor=blue_ref,citecolor=blue_ref,linkcolor=blue_ref]{hyperref}

\usepackage{comment} 
\usepackage{soul} 
\setstcolor{red}
\usepackage{cancel}
\usepackage{lipsum}

\usepackage{pgffor}

\usepackage{pdfpages} 
\makeatletter
\AtBeginDocument{\let\LS@rot\@undefined}
\makeatother

\usepackage[us,hhmmss]{datetime} 

\newcommand{\m}[1]{\mathit{#1}}
\newcommand{\f}[1]{\mathrm{#1}}
\newcommand{\map}[1]{\ifcat\noexpand#1\relax#1\else{\mathcal{#1}}\fi}
\newcommand{\set}[1]{\mathbbol{#1}}
\newcommand{\arccosh}{\mathrm{arccosh}\hspace{0.04cm}}

\newcommand{\ma}[1]{\begin{bmatrix} #1 \end{bmatrix}}
\newcommand{\eq}[2]{\begin{equation} \label{eq:#1} #2 \end{equation}}
\newcommand{\alsub}[2]{\begin{subequations}\label{eq:#1}\begin{align} #2 \end{align}\end{subequations}}
\let\geq\geqslant

\newcommand{\qs}{S}
\newcommand{\id}[1]{\mathds{1}_{#1}}
\newcommand{\hil}{\mathscr{H}}
\newcommand{\los}{\mathfrak{L}}
\newcommand{\trc}{\mathrm{tr}}
\newcommand{\ptrc}[1]{\mathrm{tr}_{\mathrm{#1}}}
\newcommand{\tra}{\intercal}

\def\HarvardSEAS{John A. Paulson School of Engineering and Applied Sciences, Harvard University, Cambridge, Massachusetts 02138, USA}

\begin{document}

\title{Information back-flow in quantum non-Markovian dynamics and its connection to teleportation}

\author{Spyros Tserkis}
\email{spyrostserkis@gmail.com}
\affiliation{\HarvardSEAS}
\author{Kade Head-Marsden}
\affiliation{\HarvardSEAS}
\author{Prineha Narang}
\email{prineha@seas.harvard.edu}
\affiliation{\HarvardSEAS}


\begin{abstract}
A quantum process is called non-Markovian when memory effects take place during its evolution. Quantum non-Markovianity is a phenomenon typically associated with the information back-flow from the environment to the principal system, however it has been shown that such an effect is not necessary. In this work, we establish a connection between quantum non-Markovianity and the protocol of quantum teleportation in both discrete and continuous-variable systems. We also show how information flows during a teleportation protocol between the principal system and the environment in a bidirectional way leading up to a state revival.
\end{abstract}

\maketitle

\section{Introduction}
Markovianity~\cite{Renyi_B_07}, in classical systems, refers to stochastic processes where the past of the system can influence its future only through its present state. Otherwise, the process is called non-Markovian, and can be interpreted as if the system contains a memory. Given that quantum mechanics is an inherently statistical theory, Markovian and non-Markovian phenomena naturally arise in quantum systems too~\cite{Wolf_Cirac_CMP_08, Rivas_Huelga_Plenio_RPP_14, Caruso_etal_RMP_14, Breuer_etal_RMP_16, deVega_Alonso_RMP_17, Li_Hall_Wiseman_PR_18, Milz_Modi_PRX_21, Ciccarello_etal_PR_22}. Even though non-Markovianity is typically associated with information flow from the principal system to the environment and back into the principal system~\cite{Breuer_Laine_Piilo_PRL_09, Piilo_etal_PRA_09, Laine_Piilo_Breuer_PRA_10}, called \textit{information back-flow}, it has been shown that this effect not only is not necessary~\cite{LoFranco_etal_PRA_12, Chruscinski_Wudarski_PLA_13, Megier_etal_SR_17}, but maximal non-Markovianity, i.e., complete state revival, can be achieved in its absence~\cite{Chruscinski_Maniscalco_PRL_14, Budini_PRA_18}.

Teleportation is a non-classical application that is conventionally described through a measurement-based process~\cite{Bennett_etal_PRL_93, Braunstein_Kimble_98}. The information flow during the teleportation protocol has been initially discussed in Refs.~\cite{Deutsch_Hayden_PRSLA_00, Horodecki_Horodecki_Horodecki_PRA_01}, but no connection was drawn to the concept of non-Markovianity. In this work, we show that the protocol of measurement-free teleportation (which is mathematically identical to the measurement-based protocol) is a time-homogeneous maximally non-Markovian process. In particular, we show this connection for both discrete-variable teleportation, which we extend from two-dimensional states~\cite{Brassard_Braunstein_Cleve_PD_98} to finite-dimensional ones, and continuous-variable states~\cite{Ralph_OL_99}. We also show how the non-Markovian nature of teleportation is entirely based on an information back-flow effect, which can be explained through the more general observation that time-homogeneity is a sufficient condition for any state revival in a non-Markovian quantum process to originate from information back-flow.

\section{Quantum Systems}

Let $\qs$ be a $d \times d$ density matrix representing a quantum state in the set of linear operators $\los[\hil_d^{(\mathrm{S})}]$, where $\hil_d^{(\mathrm{S})}$ denotes the Hilbert space~\cite{Breuer_Petruccione_B_02, Nielsen_Chuang_B_10}, representing the principal system. Due to the Stinespring dilation theorem~\cite{Stinespring_PAMS_55}, a completely-positive trace-preserving (CPTP) map $\map{M}_{(t, t_0)}: \los[\hil_d^{(\mathrm{S})}] \rightarrow \los[\hil_d^{(\mathrm{S})}]$ can be written as
\begin{equation}\label{eq:stinespring}
\map{M}_{(t, t_0)}(\qs_{t_0}) = \ptrc{E} \left[ \m{U}(t, t_0)(\qs_{t_0} \otimes \m{E}_{t_0}) \m{U}^{\dag}(t, t_0) \right] = \qs_{t},
\end{equation}
where: (i) $\m{E} \in \los[\hil_d^{(\mathrm{E})}]$ denotes a quantum state that belongs to the environment; (ii) $\m{U}(t, t_0)=\map{T} \big[e^{-i \int_{t_0}^t \dd s \m{H}(s)} \big] \in \los[\hil_d^{(\mathrm{S})}] \otimes \los[\hil_d^{(\mathrm{E})}]$ is a unitary operator associated with the global system-environment Hamiltonian, with $\map{T}$ being the chronological time-ordering superoperator; and (iii) $\ptrc{E}$ the partial trace over the environment. When $\map{M}_{(t, t_0)} = \map{M}_{\tau}$ with $\tau = t - t_0$ the quantum process is called \textit{time-homogeneous}, otherwise it is called \textit{time-inhomogeneous}~\cite{Rivas_Huelga_Plenio_PRL_10}.

The quantum evolution in Eq.~\eqref{eq:stinespring} can take the equivalent operator-sum representation~\cite{Sudarshan_Mathews_Rau_PR_61, Hellwig_Kraus_CMP_70}
\begin{equation}\label{eq:kraus_1}
\map{M}_{(t, t_0)}(\qs_{t_0}) = \sum_i \m{F}_i(t, t_0) \qs_{t_0} \m{F}_i^{\dag}(t, t_0) ,
\end{equation}
where the set of operators $\set{F}\equiv \{ \m{F}_i(t, t_0) \}$ are known as Kraus operators, satisfying $\sum_i \m{F}_{i}^{\dag}(t, t_0) \m{F}_{i}(t, t_0)= \id{}$. For a subset of CPTP maps, e.g., qubit channels, Eq.~\eqref{eq:kraus_1} can be written in terms of a probability distribution $\set{P} \equiv \{p_i(t_0, t)\}$ and a set of unitary operators $\set{U}\equiv \{ \m{U}_i(t_0, t) \}$,
\begin{equation}\label{eq:unitary_dec}
\map{M}_{(t, t_0)}(\qs_{t_0}) = \sum_i p_i(t, t_0) \m{U}_i(t, t_0) \qs_{t_0} \m{U}_i^{\dag} (t, t_0).
\end{equation}
The above CPTP maps, are referred to as mixed-unitary maps~\cite{Mendl_Wolf_CMP_09, Watrous_B_18}.

\section{Quantum Markovianity}

\begin{figure}[t]
\centering
\includegraphics[width=\linewidth]{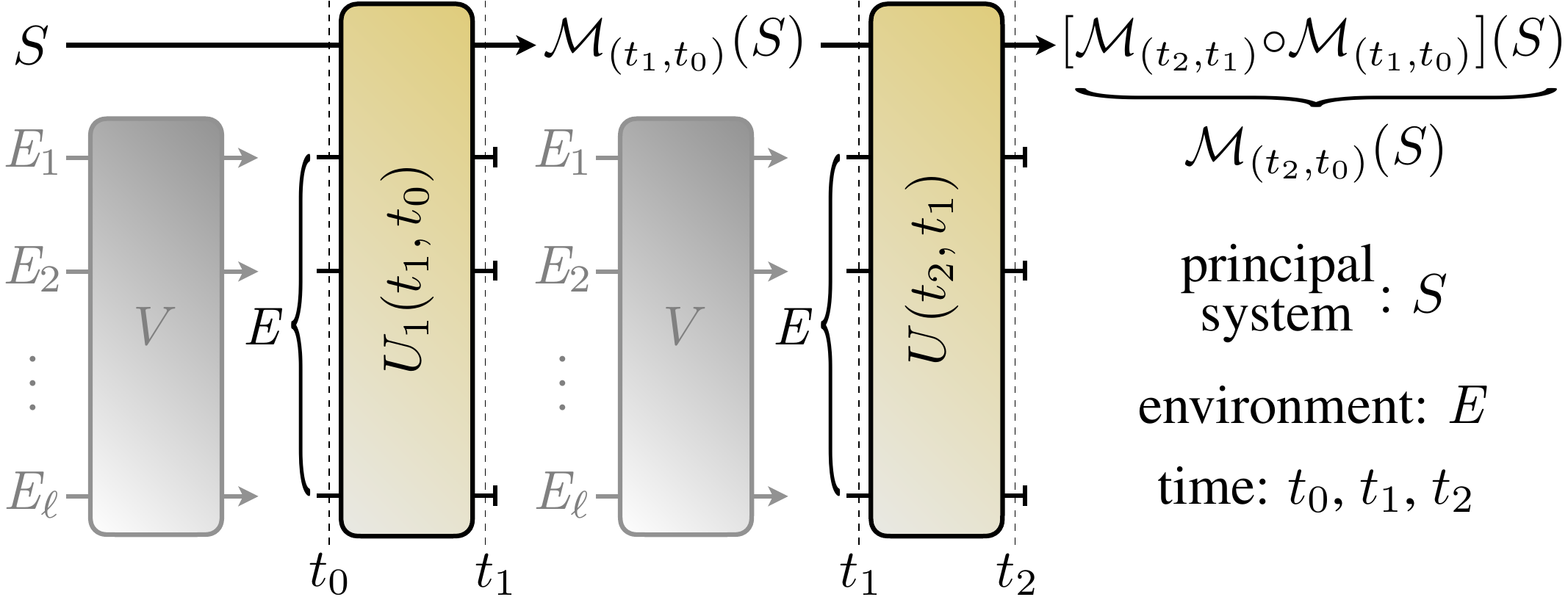}
\caption{Markovian system. During the time interval $(t_1,t_0)$ the quantum state $\qs$ is coupled to the environment, $\m{E}$, through a unitary operation $\m{U}(t_1,t_0)$. $\m{E}$ is in general an ensemble of states $\set{E}=\{ \m{E}_1,  \m{E}_2, \cdots , \m{E}_{\ell} \}$ coupled with each other through a unitary operation $\m{V}$. After this interaction all environment states are discarded, and the output state is $\map{M}_{(t_1,t_2)}(\qs)$. During the time interval $(t_2,t_1)$, the state $\map{M}_{(t_1,t_0)}(\qs)$ is coupled again with $\m{E}$ through a unitary operation $\m{U}(t_2,t_1)$, and upon discarding the interacted environment states the output final state is $[\map{M}_{(t_2,t_1)} {\circ} \map{M}_{(t_1,t_0)}](\qs)$=$\map{M}_{(t_2,t_0)}(\qs)$.}
\label{fig:markovianity}
\end{figure}

Markovianity in quantum systems admits various definitions~\cite{Wolf_Cirac_CMP_08, Rivas_Huelga_Plenio_RPP_14, Caruso_etal_RMP_14, Breuer_etal_RMP_16, deVega_Alonso_RMP_17, Li_Hall_Wiseman_PR_18, Milz_Modi_PRX_21, Ciccarello_etal_PR_22}. Here, we follow Ref.~\cite{Rivas_Huelga_Plenio_PRL_10}, and call a quantum process \textit{Markovian} if its corresponding CPTP map $\map{M}$ is divisible. A CPTP map that takes place during a time interval $(t_n,t_0)$, with $n\in \set{N}$ and $t_0 < t_n$, is called divisible if it can be written as the concatenation of the CPTP maps of each time sub-interval, i.e.,
\begin{equation}\label{eq:quantumivisibility}
\map{M}_{(t_n,t_0)}=\map{M}_{(t_n,t_{n-1})} \circ  \cdots \circ \map{M}_{(t_2,t_1)} \circ \map{M}_{(t_1,t_0)} ,
\end{equation}
with $t_0 < t_1 < \cdots < t_n$, assuming that the environment state $\m{E}$ associated to the CPTP map of each sub-interval does not carry any system-environment correlations from past interactions. The divisible process in Eq.~\eqref{eq:quantumivisibility} for $n=2$ is depicted in Fig.~\ref{fig:markovianity}, where the system is coupled to the environment through two subsequent unitary transformations and the environment states are traced out after each transformation. When a quantum process is not divisible it is called \textit{non-Markovian}. Experimentally, non-Markovianity can be simulated through different platforms, such as optical systems~\cite{Liu_etal_NP_11, Chiuri_etal_SP_12, Liu_etal_SR_13, Cialdi_etal_APL_17, Liu_etal_NC_18, Cuevas_etal_SR_19}, trapped ions~\cite{Wittemer_etal_PRA_18}, NMR~\cite{Bernardes_etal_SR_16}, and quantum computers~\cite{Sweke_etal_PRA_16, GarciaPerez_Rossi_Maniscalco_NPJ_20, Hu_Xia_Kais_SR_20, HeadMarsden_etal_PRR_21}. Non-Markovianity has also been explored under scenarios where two channels exhibit indefinite causal order~\cite{Cheong_Pradana_Chew_PRA_24, Maity_Bhattacharya_JPA_24}.

There are multiple ways to identify non-Markovianity in a quantum system, one of them being through checking the evolution of fidelity between two states. Fidelity is a measure used to distinguish two quantum states, defined as $\f{F}(\varPsi_i,\varPsi_j) \coloneqq \big( \trc \sqrt{\smash[t]{\sqrt{\varPsi_i}} \smash[b]{\varPsi_j} \smash[b]{\sqrt{\varPsi_i}}} \big)^2$~\cite{Jozsa_JMO_94}, with $\f{F}(\varPsi_i,\varPsi_j) =1 \Leftrightarrow \varPsi_i = \varPsi_j$. This distinguishability measure is monotonic under complete positive maps, i.e., $ \f{F}[ \map{M}(\varPsi_i), \map{M}(\varPsi_j)] \geq  \f{F} ( \varPsi_i, \varPsi_j )$, and thus a negation of this condition implies non-Markovianity~\cite{Rivas_Huelga_Plenio_RPP_14}.

\section{Information back-flow effect}

A revival of a state in the principal system during its evolution through a quantum process is a sufficient (but not necessary) condition to detect non-Markovianity~\cite{Rivas_Huelga_Plenio_RPP_14}, which can be checked through the non-monotonous behavior of the distinguishability between two evolving quantum states. In Refs.~\cite{Breuer_Laine_Piilo_PRL_09, Piilo_etal_PRA_09, Laine_Piilo_Breuer_PRA_10} this non-monotonous behavior of distinguishability was attributed  to the back-flow of information from the environment to the principal system, however it was later shown~\cite{LoFranco_etal_PRA_12, Chruscinski_Wudarski_PLA_13, Megier_etal_SR_17, Chruscinski_Maniscalco_PRL_14, Budini_PRA_18} that the two concepts are not always equivalent. In particular, a time-inhomogeneous non-Markovian process that evolves according to Eq.~\eqref{eq:unitary_dec} can revive a quantum state without any information back-flow effect~\footnote{Note that here by ``information'' we refer to quantum information, and not classical information measured by an agent outside of the system as in Ref.~\cite{Buscemi_Datta_PRA_16}.}. This is possible since in time-inhomogeneous processes the interaction between the system and the environment evolves over time, and thus this time-dependency must originate from an external source, i.e., the entire system under question is not closed. On the other hand, in time-homogeneous processes the interaction does not evolve over time, so the total system is closed~\cite{Breuer_Petruccione_B_02, Nielsen_Chuang_B_10}, which implies that any state revival must originate from information back-flow due to the conservation of quantum information principle~\cite{Horodecki_Horodecki_PLA_98}. It is worth noting that information back-flow in a positive trace-preserving (PTP) map $\map{M}$ is equivalent its indivisibility~\cite{PhysRevA.92.042108}.

Below, we show that, at least under the conditions we studied, the teleportation protocol corresponds to a maximally non-Markovian time-homogeneous quantum process.

\section{Teleportation}

In general, any teleportation protocol involves two agents, called Alice and Bob. Alice has an input state $\qs$ that she wants to send to Bob. The two agents also share a bipartite entangled state $\m{R}$, known as the resource state, and a perfect teleportation requires a maximally entangled resource state, given by $\m{R}=\varPhi=\ketbra{\phi}{\phi}$, with $\ket{\phi}=\frac{1}{\sqrt{d}}\sum_{i=0}^{d-1} \ket{ii}$, where $d$ denotes the dimensions of the Hilbert space, and $\{ \ket{i}\}_{i=0}^{d-1}$ an orthonormal basis. Note that $\ket{ii} \equiv \ket{i} {\otimes} \ket{i}$.

Let us consider the measurement-free teleportation protocol~\cite{Brassard_Braunstein_Cleve_PD_98, Ralph_OL_99} which is a modified version of the conventional (measurement-based) teleportation~\cite{Bennett_etal_PRL_93, Braunstein_Kimble_98}. This protocol can be summarized as the process where at Alice's side the input state $\qs$ is coupled to an entangled state $\m{R}$. The outcome of this interaction is sent to Bob coherently, where the input state $\qs$ is recovered via another interaction.

\subsection{Discrete-variable Teleportation}
\label{sec:dv_teleportation}

\begin{figure}[t]
\centering
\includegraphics[width=\linewidth]{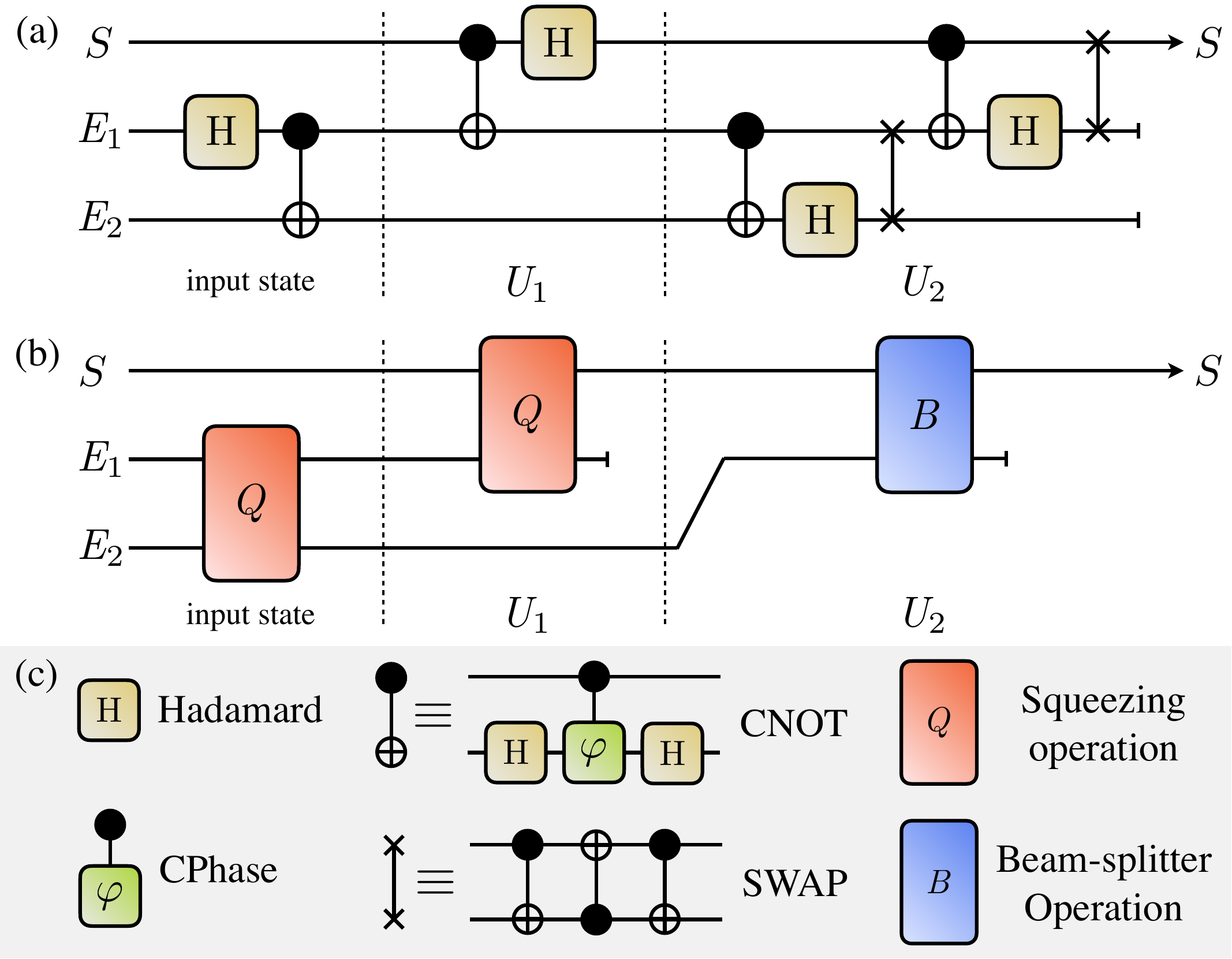}
\caption{Measurement-free teleportation circuits. In panel (a) the DV teleportation is presented, and in panel (b) the CV one. The dashed lines in each panel denote the three subsequent operations of the protocols, $\m{U}_1$, $\m{U}_2$, and $\m{U}_3$. In panel (c) the gates and operations used in the circuits are labeled. The specific parameters used in the gate and operations above are given in Eq.~\eqref{eq:op_dv} and Eq.~\eqref{eq:op_cv}.}
\label{fig:teleportation}
\end{figure}

Measurement-free teleportation in discrete-variable (DV) systems was introduced by Brassard, Braunstein, and Cleve (BBC) for 2-dimensional states, i.e., qubits~\cite{Brassard_Braunstein_Cleve_PD_98} and it was experimentally realized for the first time in Ref.~\cite{Nielsen_Knill_Laflamme_N_98}. The BBC protocol requires the use of two quantum gates: (i) the Hadamard and (ii) the controlled-NOT (CNOT). For the purposes of this work, the SWAP gate is also used so that the final state appears in the principal system. Note that the Hilbert space on which the recovered state appears is irrelevant since Bob has access to all of them. This modified version of the BCC protocol is depicted in Fig.~\ref{fig:teleportation}~(a).

Here, we extend the BBC protocol in arbitrary finite $d$-dimensional states, i.e., qudits. In order to do so, the aforementioned quantum gates need to be appropriately generalized. The generalized Hadamard gate corresponds to the discrete Fourier transform $\mathrm{H}(\theta) \ket{j} \coloneqq \frac{1}{\sqrt{d}} \sum_{i=0}^{d-1} e^{\frac{ij \theta \imath}{d}} \ket{i}$, where $\{ \ket{i} \}_{i=0}^{d-1}$ and $\{ \ket{j} \}_{j=0}^{d-1}$ are orthonormal bases, $\theta \in [0, 2d \pi ]$, and $\imath \coloneqq \sqrt{-1}$. The generalized CNOT gate can be defined as $\mathrm{CNOT}(\varphi, \theta) \ket{ij} \coloneqq [\id{} {\otimes} \mathrm{H}(\theta)] \mathrm{CPhase}(\varphi) [\id{} {\otimes} \mathrm{H}(\theta)] \ket{ij}$\cite{GarciaEscartin_ChamorroPosada_QIP_13}, where $\mathrm{CPhase}(\varphi) \ket{ij} \coloneqq e^{\frac{i j \varphi \imath}{d}} \ket{ij}$ is the generalized controlled-phase (CPhase) gate with $\varphi \in [0, 2d \pi ]$~\footnote{CNOT gate admits multiple generalizations~\cite{Wang_etal_FP_20}, see for example Ref.~\cite{Alber_etal_JPAMG_01}.}. Finally, the generalized SWAP gate, $\mathrm{SWAP}(\varphi, \theta)$, is constructed through a sequence of three generalized CNOT gates where the middle one is inverted~\cite{GarciaEscartin_ChamorroPosada_QIP_13}, as shown in Fig.~\ref{fig:teleportation}~(c).

Let $\qs $ be an arbitrary quantum state belonging to Alice, and $\m{E}$ the state that corresponds to to the environment. Note that by ``environment'' here we refer to the system that we trace out after the interaction takes place. The DV measurement-free teleportation can be realized through the following non-divisible operation
\begin{equation}
\ptrc{E_{1,2}} \left[ \m{U}_2\m{U}_1 (\qs \otimes \m{E}) \m{U}^{\dag}_1\m{U}^{\dag}_2  \right] = \qs,
\end{equation}
where
\begin{subequations} \label{eq:op_dv}
\begin{align}
\m{U}_1&=[\mathrm{H}(2 \pi) \otimes \id{} \otimes \id{}][\mathrm{CNOT}(2 \pi, 2d\pi{-}2\pi) \otimes \id{}], \\
\m{U}_2&=[\mathrm{SWAP}(2 \pi, 2 \pi) \otimes \id{}]  \nonumber \\
& \hspace{0.46cm} {\times} [\id{} \otimes \mathrm{H}(2 \pi) \otimes \id{}] [\mathrm{CNOT}(2 \pi, 2 \pi) \otimes \id{}]   \nonumber \\
& \hspace{0.46cm} {\times} [\id{} \otimes \mathrm{SWAP}(2 \pi, 2 \pi)]  \nonumber \\
& \hspace{0.46cm} {\times} [\id{} \otimes \id{} \otimes \mathrm{H}(2d\pi{-}2\pi)] [\id{} \otimes \mathrm{CNOT}(2 \pi, 2 \pi)] .
\end{align}
\end{subequations}

In order to see how information flows from one system to another, let us consider for simplicity a pure state $\ket{s}$ for the principal system and the maximally entangled state $\ket{\phi}$ for the environment.

The operation $\m{U}_1$ corresponds to the coupling between the principal state and the environment, that effectively leads to information about the state $\ket{s}$ moving from S to $\mathrm{E}_2$, described by the identity~\cite{Braunstein_PRA_96, Roa_Delgado_Fuentes_PRA_03},
\begin{equation}
\m{U}_1 ( \ket{s} \otimes \ket{\phi} ) = \frac{1}{d}\sum_{i,j=0}^{d-1} \ket{ij} \otimes \m{X}_d^{d-j} \m{Z}_d^{i} \ket{s} ,
\end{equation}
where $\m{X}_d= \sum_{i=0}^{d-1} \ketbra{i\oplus1}{i}$ and $\m{Z}_d= \sum_{i=0}^{d-1} e^{\frac{i2\pi \imath}{d}} \ketbra{i}{i}$ are two unitary operations, with $\oplus$ denoting the modulo-$d$ addition when it is applied on scalar values. The final stage, $\m{U}_2$, brings the state $\ket{s}$ back to the principal system, 
\begin{equation}\label{eq:dv_revival}
\m{U}_2 \left( \frac{1}{d}\sum_{i,j=0}^{d-1} \ket{ij} \otimes \m{X}_d^{d-j} \m{Z}_d^{i} \ket{s} \right) = \ket{s} \otimes \ket{++} ,
\end{equation}
where $\ket{+}=\frac{1}{\sqrt{d}}\sum_{i=0}^{d-1} \ket{i}$.

\begin{figure*}[t]
\centering
\includegraphics[width=\textwidth]{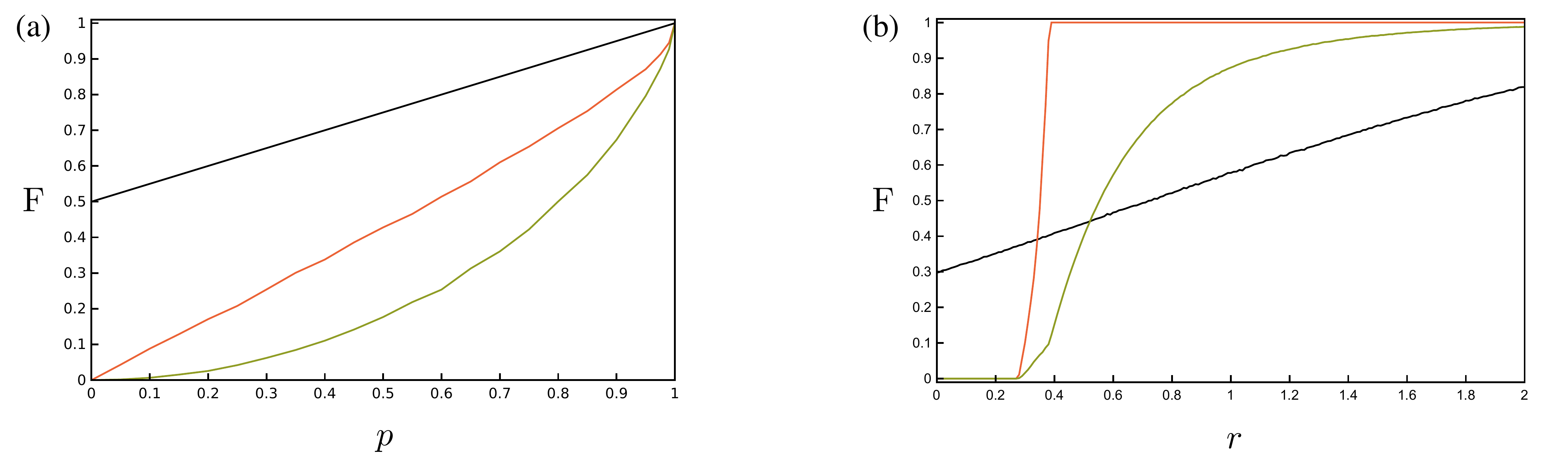}
\caption{Panel (a) corresponds to the DV teleportation protocol. The black curve corresponds to the fidelity between the input and the output state averaged over the entire sample of random inputs. The increasing trend indicates that the less noisy is the resource state the more successful is the teleportation. At $p=1$, which corresponds to a maximally entangled state for the resource state, we have full state revival. The red curve corresponds to the ratio of the input random pair of states that exhibit non-Markovianity, for which we also calculate the revival of fidelity, i.e., the decrease of fidelity between the first and second step over the initial increase of fidelity from the beginning to the first step, depicted with the green curve. Panel (b) corresponds to the CV teleportation protocol. Analogously to panel (a) the black curve corresponds to the input-output fidelity, the red curve to the ratio of the states that exhibit non-Markovianity, and the green curve to the revival of fidelity.}
\label{fig:fidelity}
\end{figure*}

In order to examine the non-Markovian nature of teleportation, we run a simulation for $10^6$ randomly created input pairs of pure states, $\qs$ and $\tilde{\qs}$, and (separately) propagate them through the teleportation circuit in Fig.~\ref{fig:teleportation}~(a). As a resource state we consider the Werner state, $W = p \m{\Phi} + (1-p) \id{}/4$, where $p$ is the error-free probability. We are interested in the evolution of states at three distinct points in the circuit: (i) in the beginning, (ii) after operation $U_1$, and after operation $U_2$.

The black curve in Fig.~\ref{fig:fidelity}~(a) corresponds to the fidelity between the input and output state, $\f{F}(\qs, \qs_2)$, averaged over all propagated states. As expected, the higher the value of $p$, i.e., the closer the Werner state is to the maximally entangled state, the larger the fidelity. The red curve corresponds to the ratio of the pairs of the sample for which non-Markovianity was identified through a non-monotonic behavior of the fidelity between the first and the second step, i.e., $\f{F}(\qs_2, \tilde{\qs}_2) < \f{F}(\qs_1, \tilde{\qs}_1) $. Note that several quantifiers exist for non-Markovianity~\cite{Rivas_Huelga_Plenio_RPP_14, Addis_etal_PRA_14} and thus any valid measure of them provides qualitatively the same behavior. The green curve corresponds to the average revival of fidelity when non-Markovianity is observed, computed as $\frac{\f{F}(\qs_1, \tilde{\qs}_1) - \f{F}(\qs_2, \tilde{\qs}_2)}{\f{F}(\qs_1, \tilde{\qs}_1)-\f{F}(\qs, \tilde{\qs})} $. We observe that the less noisy the system the bigger the effect of non-Markovianity, and the point that all states exhibit non-Markovianity is the same point where a complete revival of the state is observed. Interestingly, non-Markovianity appears throughout the range of $p$ even though for $p<1/3$ the resource state is not entangled, meaning that non-Markovianity is closer connected to non-classical correlations, e.g., quantum discord.

\subsection{Continuous-variable Teleportation}

Measurement-free teleportation in continuous-variable (CV) systems~\cite{Serafini_B_17} was proposed by Ralph~\cite{Ralph_OL_99} by the name ``all-optical teleportation'' (see also Ref.~\cite{Tserkis_Hosseinidehaj_Ralph_PRR_20}) and it was experimentally realized in Ref.~\cite{Liu_Lou_Jing_NC_20}. This protocol can be realized through the following non-divisible operation
\begin{equation}
\ptrc{E_2} \left\{ \m{U}_2 \, \ptrc{E_1} [ \m{U}_1(\qs \otimes \m{E}) \m{U}^{\dag}_1 ] \m{U}^{\dag}_2 \right\} = \qs,
\end{equation}
where $\qs$ belongs to Alice and $\m{E}$ to the environment (similarly to the DV case the term ``environment'' is used to label the modes that will be traced out after the interaction), as seen in Fig.~\ref{fig:teleportation}~(b). The unitary operations are given below
\begin{subequations} \label{eq:op_cv}
\begin{align}
\m{U}_1&= \m{Q}[\arccosh (\sqrt{g})] \otimes \id{\infty} , \\
\m{U}_2&= \m{B}(1/g) .
\end{align}
\end{subequations}
$\m{Q}(r)$ is the two-mode squeezing operation $\m{Q}(r) \coloneqq \exp[r(\m{A}_1\m{A}_2-\m{A}^{\dag}_1\m{A}^{\dag}_2)/2]$ with $\m{A}$ denoting the annihilation operator and $r \in \set{R}$ the real-valued squeezing parameter. The subscripts of the operators refer to the mode on which they are applied. The two-mode squeezing operation is the entanglement operation in CV states, i.e., $\m{Q}(r) \ket{00} = \sqrt{1- \tanh^2 r} \sum_{i=0}^{\infty} (-\tanh r)^i \ket{ii}$, known as the two-mode squeezed vacuum, which becomes maximally entangled in the limit of $r \rightarrow \infty$. $\m{B}(\tau)$ is the beam-splitter operation $\m{B}(\tau) \coloneqq \exp[\tau(\m{A}_1^{\dag} \m{A}_2-\m{A}_1 \m{A}_2^{\dag})] $ with $\tau \in [0, 1]$ being the transmissivity parameter.

Regarding information flow in this protocol, consider for simplicity a Gaussian CV system~\cite{Weedbrook_etal_RMP_12}, where the quantum states can be fully represented through covariance matrices, $\m{V}$, and the unitary operations through symplectic transformations, $\varSigma$, i.e., $\m{V} \rightarrow \varSigma \m{V} \varSigma^{\tra}$. Let us have the input state $\m{V}_S$ and a two-mode squeezed vacuum $\m{V}_{\varPhi}$.

The first stage involves the symplectic transformation $\varSigma_1$ and a partial trace over the first mode of the environment, so the covariance matrix transforms into
\begin{equation}\label{eq:cv_s2}
\ptrc{E_1} \left[ \varSigma_1 \left( \m{V}_S \oplus \m{V}_{\varPhi} \right) \varSigma_1^{\tra} \right] .
\end{equation}

Given that prior to the partial trace the matrix in Eq.~\eqref{eq:cv_s2} represents a tri-partite entangled state (see Ref.~\cite{Teh_Reid_PRA_14} for appropriate entanglement criteria), information about the input system is transferred in the form of correlations in both modes of environment. Finally, the second symplectic transformation $\varSigma_2$ results in a state revival for the principal system,
\begin{equation}\label{eq:cv_s3}
\ptrc{E_{2}} \left\{ \varSigma_2 \ptrc{E_1} \left[ \varSigma_1 \left( \m{V}_S \oplus \m{V}_{\varPhi} \right) \varSigma_1^{\tra} \right] \varSigma_2^{\tra}\right\}= \m{V}_S .
\end{equation}
More information about Eqs.~\eqref{eq:cv_s2}-\eqref{eq:cv_s3} are given in the Appendix.

In order to show the non-Markovian nature of this protocol, we follow a similar path to the DV case, by propagating $10^6$ randomly created input pairs of pure (rotated and squeezed) Gaussian states, $\m{V}$ and $\tilde{\m{V}}$, through the CV teleportation circuit. The input and output average fidelity\footnote{Fidelity between two single-mode (non-displaced) Gaussian states is defined as $\f{F}(\m{V}_i,\m{V}_j) \coloneqq 2/(\sqrt{\mu + \nu} - \sqrt{\nu})$, where $\mu = \det (\m{V}_i + \m{V}_j)$ and $\nu = (\det \m{V}_i-1)(\det\m{V}_j -1)$~\cite{Nha_Carmichael_PRA_05, Banchi_Braunstein_Pirandola_PRL_15}.} is denoted with the black curve, $\f{F}(\m{V}, \m{V}_2)$, and the ratio of states that exhibit non-Markovianity, i.e., $\f{F}(\m{V}_2,  \tilde{\m{V}}_2) < \f{F}(\m{V}_1,  \tilde{\m{V}}_1)$, is depicted with the red curve in Fig.~\ref{fig:fidelity}~(b). With the green curve we have the average revival of fidelity $\frac{\f{F}(V_1, \tilde{V}_1) - \f{F}(V_2, \tilde{V}_2)}{\f{F}(V_1, \tilde{V}_1)-\f{F}(V, \tilde{V})} $ for the cases where non-Markovianity is identified. As a resource state we consider a two-mode squeezed vacuum $\m{Q}(r) \ket{00}$ with $r \in [0,2]$\footnote{The selected range of the squeezing parameter is accessible with current technology, since $r_{\mathrm{max}}=2$ corresponds to $\sim$17.4dB of squeezing.}. Similarly to the DV case, we observe that the fidelity between the input and output states increases with $r$ but it does not reach unity since the CV measurement-free teleportation works perfectly in the limit of  $r \rightarrow \infty$ and $g \rightarrow \infty$. Interestingly, non-Markovianity does not kick off immediately, but it requires squeezing of about $r\sim0.25$. Also, at about $r \sim 0.4$ all of the input states exhibit non-Markovianity. Note that the resource state is entangled for any non-zero squeezing parameter suggesting that the non-Markovian character of the CV teleportation is not directly related to entanglement.

It should be noted that the DV and the CV teleportations are not physically the same protocol represented in different dimensions, but two distinct protocols that can revive a quantum state, so there is no reason to expect them to have the same non-Markovian characteristics. However, both the DV and the CV case agree on the existence (or not) of non-Markovianity in the extreme cases for their corresponding studied resource state.

\section{Conclusion}

In this work it is shown that when it is seen from an open quantum system point of view, the protocol of teleportation is a non-Markovian quantum operation. This connection is achieved when the teleportation is studied under its measurement-free approach, which is mathematically equivalent to the conventional measurement-based one. In particular, for the continuous-variable case we employ the all-optical teleportation~\cite{Ralph_OL_99}, and for the discrete-variable case we extend to finite-dimensional states a measurement-free protocol~\cite{Brassard_Braunstein_Cleve_PD_98} that was limited to qubits. It is also discussed how the state revival is due to an information back-flow effect given the time-homogeneous nature of the process. Teleportation is a fundamental protocol in quantum information~\cite{Pirandola_etal_NP_15} that provides a building block for broader applications such as the quantum internet~\cite{Kimble_N_08, Perseguers_etal_RPP_13, Wehner_Elkouss_Hanson_S_18} and photonic quantum computers~\cite{Kok_etal_RMP_07, Bourassa_etal_Q_21, Bartolucci_etal_Q_21}. Thus, the relationship between this protocol and the phenomenon of non-Markovianity creates a new perspective on the information flow during those applications.

\textit{Note added:} Ref.~\cite{McAleese_Paternostro_arxiv_24} has recently followed up on our work, providing a deeper analysis of Section~\ref{sec:dv_teleportation}. The authors show that the connection between teleportation and non-Markovianity can be revealed by employing a more fine-grained discretization of the circuit in Fig.~\ref{fig:teleportation}(a).

\section*{Acknowledgments}
This is supported by the NSF Engineering Research Center (ERC) Center for Quantum Networks and NSF RAISE-QAC-QSA, Grant No. DMR-2037783 on ``Open Quantum Systems on Noisy Intermediate-Scale Quantum Devices''. K.H.M. is partially supported by the Department of Energy, Office of Basic Energy Sciences Grant DE-SC0019215 on ``Quantum Computing Algorithms and Applications for Coherent and Strongly Correlated Chemical Systems''. P.N. acknowledges support as a Moore Inventor Fellow through Grant No. GBMF8048 and gratefully acknowledges support from the Gordon and Betty Moore Foundation as well as support from a NSF CAREER Award under Grant No. NSF-ECCS-1944085.

\appendix
\section{Symplectic Transformations in CV Measurement-Free Teleportation}
\label{appA}

In the main paper, the information flow for the CV measurement-free teleportation is analyzed for the case of a single-mode Gaussian state~\cite{, Weedbrook_etal_RMP_12, Serafini_B_17}. A Gaussian state can be fully described by its covariance matrix $\m{V}$ given by
\eq{}{
\m{V} = \ma{
\langle \m{X}^2 \rangle & \frac{1}{2}\langle \{ \m{X}, \m{P} \} \rangle \\
\frac{1}{2}\langle \{ \m{P}, \m{X} \} \rangle  & \langle \m{P}^2 \rangle
\end{bmatrix} - \begin{bmatrix}
\langle \m{X} \rangle^2 & \langle \m{X} \rangle \langle \m{P} \rangle  \\
\langle \m{P} \rangle \langle \m{X} \rangle & \langle \m{P} \rangle^2
} ,
}
where $\m{X}\coloneqq \m{A}+\m{A}^{\dag}$ and $\m{P} \coloneqq \imath(\m{A}^{\dag}-\m{A})$ are the quadrature operators, typically called position and momentum, respectively. The unitary evolution of a Gaussian state takes the form of a symplectic transformation $\m{\Sigma}$ on the covariance matrix,
\eq{}{
\m{V} \rightarrow \m{\Sigma} \m{V} \m{\Sigma}^{\tra} .
} 
A matrix is called symplectic when $\m{\Sigma} \m{\Omega} \m{\Sigma}^{\tra}= \m{\Omega}$ where $ \m{\Omega} \coloneqq \bigoplus_{i=1}^n \omega$ with $\omega=\ma{0 & 1\\ -1& 0}$, and $n$ denotes the number of modes.

The unitary operations $\m{U}_1$ and $\m{U}_2$ given in Eq.~(11) in the main text correspond to the following symplectic transformations:
\alsub{}{
\m{\Sigma}_1 &=\ma{
\sqrt{g} & 0 & \sqrt{g-1} & 0 \\ 
0 & \sqrt{g} & 0 & -\sqrt{g-1} \\
\sqrt{g-1} & 0 & \sqrt{g} & 0 \\
0 & -\sqrt{g-1} & 0 & \sqrt{g}
}\oplus \ma{1 & 0\\ 0& 1} ,  \\
\m{\Sigma}_2 &=\ma{
\sqrt{\tau} & 0 & -\sqrt{1-\tau} & 0 \\ 
0 & \sqrt{\tau} & 0 & -\sqrt{1-\tau} \\
\sqrt{1-\tau} & 0 & \sqrt{\tau} & 0 \\
0 & \sqrt{1-\tau} & 0 & \sqrt{\tau}
} .
}

The covariance matrix for the two-mode squeezed vacuum state $\m{Q}(r) \ket{00}$ is given by
\eq{}{
\m{V}_{\m{Q}(r) \ket{00}}= \ma{
\cosh 2r & 0 & \sinh 2r & 0 \\ 
0 & \cosh 2r & 0 & -\sinh 2r \\
\sinh 2r & 0 & \cosh 2r & 0 \\
0 & -\sinh 2r & 0 & \cosh 2r
},
}
so for the maximally entangled state we have $\m{V}_{\m{\Phi}}=\lim_{r \rightarrow \infty} \m{V}_{\m{Q}(r) \ket{00}}$.

\bibliography{bibliography}

\end{document}